\documentclass[journal]{IEEEtran}
\usepackage{color}
\usepackage{balance}
\usepackage{graphicx}
\usepackage{enumerate}
\usepackage{subfigure}
\usepackage{algorithm}
\usepackage{fancyhdr}
\usepackage{algorithmic}
\usepackage{booktabs}
\usepackage{cite}
\usepackage{url}
\usepackage{amssymb}
\usepackage{amsmath}
\usepackage{float}
\usepackage{float}
\usepackage{bm}
\usepackage{amsmath,epsfig,amssymb,subfigure,bm,dsfont}
\usepackage{lipsum} 

\begin{document}
\title{Deep Reinforcement Learning Based Massive Access Management for Ultra-Reliable Low-Latency Communications}

\author{

  Helin~Yang,~\IEEEmembership{Student Member,~IEEE,}
    Zehui~Xiong,~\IEEEmembership{Student Member,~IEEE,}
        Jun~Zhao,~\IEEEmembership{Member,~IEEE,}
   Dusit~Niyato,~\IEEEmembership{Fellow,~IEEE,}
     Chau Yuen,~\IEEEmembership{Senior Member,~IEEE,}
 and Ruilong~Deng,~\IEEEmembership{Senior Member,~IEEE}

\thanks{This research is supported by Nanyang Technological University (NTU) Startup Grant,  Alibaba-NTU Singapore Joint Research Institute (JRI), Singapore Ministry of Education Academic Research Fund Tier 1 RG128/18, Tier 1 RG115/19, Tier 1 RT07/19, Tier 1 RT01/19, and Tier 2 MOE2019-T2-1-176,  NTU-WASP Joint Project, Singapore National Research Foundation under its Strategic Capability Research Centres Funding Initiative: Strategic Centre for Research in Privacy-Preserving Technologies \& Systems,  Energy Research Institute @NTU, Singapore NRF National Satellite of Excellence, Design Science and Technology for Secure Critical Infrastructure NSoE DeST-SCI2019-0012, AI Singapore 100 Experiments (100E) programme,  NTU Project for Large Vertical Take-Off \& Landing Research Platform, National Research Foundation (NRF), Singapore, under Singapore Energy Market Authority (EMA), Energy Resilience, NRF2017EWT-EP003-041, Singapore NRF2015-NRF-ISF001-2277, Singapore NRF National Satellite of Excellence, Design Science and Technology for Secure Critical Infrastructure NSoE DeST-SCI2019-0007, A*STAR-NTU-SUTD Joint Research Grant on Artificial Intelligence for the Future of Manufacturing RGANS1906, Wallenberg AI, Autonomous Systems and Software Program and Nanyang Technological University (WASP/NTU) under grant M4082187 (4080), Singapore Ministry of Education (MOE) Tier 1 (RG16/20), NTU-WeBank Joint Research Institute (JRI) (NWJ-2020-004), and Alibaba Group through Alibaba Innovative Research (AIR) Program.  \emph{ (Corresponding author: Jun Zhao.)}}

\thanks{ \small{H. Yang, Z. Xiong, J. Zhao, and D. Niyato are with the School of Computer Science and Engineering, Nanyang Technological University, Singapore 639798 (e-mail: hyang013@e.ntu.edu.sg, zxiong002@e.ntu.edu.sg, junzhao@ntu.edu.sg, dniyato@ntu.edu.sg).}}
\thanks{ \small{C. Yuen is with Singapore University of Technology and Design, Singapore 138682 (e-mail: yuenchau@sutd.edu.sg).}}

\thanks{\small{R. Deng is with the College of Control Science and Engineering, Zhejiang University, Hangzhou 310027, China, where he is also with the School of Cyber Science and Technology (e-mail: dengruilong@zju.edu.cn).}}

}


\maketitle

	 \thispagestyle{fancy}
\pagestyle{fancy}
\lhead{This paper appears in \textbf{IEEE Transactions on Wireless Communications}. \hfill \thepage \\ Please feel free to contact us for questions or remarks. \hfill \url{http://JunZhao.info/} }
\cfoot{}
\renewcommand{\headrulewidth}{0.4pt}
\renewcommand{\footrulewidth}{0pt}

\begin{abstract}

With the rapid deployment of the Internet of Things (IoT), fifth-generation (5G) and beyond 5G networks are required to support massive access of a huge number of devices over limited radio spectrum radio. In wireless networks, different devices have various quality-of-service (QoS) requirements, ranging from ultra-reliable low latency communications (URLLC) to high transmission data rates. In this context, we present a joint energy-efficient subchannel assignment and power control approach to manage massive access requests while maximizing network energy efficiency (EE) and guaranteeing different QoS requirements. The latency constraint is transformed into a data rate constraint which makes the optimization problem tractable before modelling it as a multi-agent reinforcement learning problem. A distributed cooperative massive access approach based on deep reinforcement learning (DRL) is proposed to address the problem while meeting both reliability and latency constraints on URLLC services in massive access scenario. In addition, transfer learning and cooperative learning mechanisms are employed  to enable communication links to work cooperatively in a distributed manner,  which enhances  the network performance and access success probability. Simulation results clearly show that the proposed distributed cooperative learning approach outperforms other existing approaches in terms of meeting EE and improving the transmission success probability in massive access scenario. 

\end{abstract}

\IEEEpeerreviewmaketitle
\section{Introduction}

\IEEEPARstart{W}{ith} the rapid development of the Internet of Things (IoT), more and more IoT devices access wireless networks to support diverse applications, e.g., smart city, intelligent transportation, smart industry and healthcare (eHealth) [1], [2]. It can be predicted that the number of IoT devices will exceed 20 billion in 2020 and reach hundreds of billions in 2030 [1]. In this context, the upcoming fifth generation (5G) and beyond 5G (B5G) networks are required to provide seamless access and diverse services for massive IoT devices. Due to massive access of a large number of devices over limited radio spectrum, the deluge of spectrum access requests may lead to severe congestion with low transmission success probability [1], [3].  Considering the explosive increase in the number of devices, it is essential to improve the access efficiency in 5G networks for accommodating massive access with various quality-of-service (QoS) guarantees. For QoS guarantee, in 5G networks, ultra-reliable and low latency communications (URLLC) is one of the most challenging services with stringent low latency and high reliability requirements, i.e., in 3GPP, a general URLLC requirement of a one-way radio is 99.999\% target reliability with 1 ms latency [4]. Consequently, URLLC entails great difficulty in massive access in 5G and B5G wireless networks.

To relieve the radio access network congestion resulted from massive access, one of the simplest spectrum access schemes, termed random access procedure, was widely investigated recently. So far, there has been lots of research on massive random access for massive machine-type communications (mMTC), IoT networks and machine-to-machine (M2M) communications  [5]-[12]. The authors in [5] and [6] proposed their contention-based random access models to enhance the access success probability and reduce the transmission delay. In [7], a two-stage random-access-based massive IoT uplink transmission protocol was presented to deal with the congestion caused by mMTC devices. Liu $et~al.$ in [8] investigated a priority-based multiple access  protocol to ensure the fairness of different devices. Furthermore, grant-based and grant-free are two common random access schemes which can provide devices’ access statuses after processing device detection and channel estimation [3], [9], [10],  but accurate channel state information (CSI) requirements of a massive number of devices may be impractical.  Besides, several methods  were presented to enhance the traditional random access performance, such as access class barring (ACB), slotted access, and backoff [11], [12]. For instance, in [12], an efficient random access procedure based on ACB was investigated to decrease the access delay and the power consumption when wireless networks have a congestion problem resulted from massive access.

The aforementioned approaches [5]-[12] based on random access are simple, flexible and could be applied without central coordinator with massive wireless connections. However, these proposals achieve limited improvement and high access failures which remain the performance bottleneck for massive access. In particular, the high transmission success probability is not easily guaranteed when the devices have strict URLLC requirements.  To satisfy the critical requirements of URLLC in massive IoT or mMTC, many studies have presented the advanced spectrum access schemes [13]-[19]. Weerasinghe $et~al.$ proposed a priority-based massive access approach to support reliable and low latency access for mMTC devices [13], where devices are categorized into  a number of groups with different priority access levels. A probability density function of signal-to-noise ratio (SNR) was derived for a large number of uplink URLLC devices in [14], and numerical results verified that the presented model can satisfy the critical requirements of URLLC. Popovski $et~al.$ in [15] discussed the principles of wireless access for URLLC and provided a perspective on the relationship between latency, packet size and bandwidth. In [16] and [17], grant-free spectrum access was adopted to reduce transmission latency as well as improve the spectrum utilization in URLLC scenario. In [18] and [19], different resource management schemes were developed  to shown how to update the system parameters that allow meeting the URLLC requirements in industrial IoT networks, since industrial automation requires strict low latency and high reliability for manufacturing control. Nevertheless, only a few literatures [13] and [14] investigated how to meet strict URRLC requirements in massive access scenario, and the optimization objective of these two studies are just a single time slot optimization problem, where the massive access decision approaches may converge to the sub-optimal solution and obtain the greedy-search like performance due to the ignorance of the historical network state and the long term benefit.

\subsection{Related Works}

Recently, several emerging technologies of 5G, i.e., massive multiple-input multiple-output (MIMO), non-orthogonal multiple access (NOMA) and device-to-device (D2D) communications are applied to support massive connectivity over limited available radio resources. Chen $et~al.$ in [12] and [20] presented non-orthogonal communication frameworks based on massive NOMA to support massive connections, and the transmit power values were optimized to mitigate severe co-channel interference by using interference cancellation techniques [21]. In addition, an application-specific NOMA-based communication architecture was investigated for future URLLC Tactile Internet [22]. In [14], [15], [23] and [24], the authors presented coordinated and uncoordinated access protocols to support massive connectivity in massive MIMO systems by exploiting large spatial degrees of freedom to admit massive IoT devices. Specifically,  the authors in [14] and [15] discussed that the massive MIMO system can be acted as a natural enabler for URLLC where multiple antenna systems can support high capacity, spatial multiplexing and diversity links. Moreover, a potential solution for the massive access congestion problem is to offload the large amount of traffic onto D2D communication links [25], [26], which can directly reduce devices' energy consumption and transmission delay, as well as improve spectrum efficiency. D2D-based transmission protocols for supporting URLLC services were proposed in [27] and [28], where devices are classified into a number of groups based on their QoS requirements, i.e., stringent low latency and high reliability requirements, with radio resources allocated accordingly.

In addition, energy efficiency (EE) plays an important role in green wireless networks. The reasons are that most of devices (e.g., sensors, actuators and wearable devices) are power constrained and energy consumption is massive and expensive under high-density scenario of devices. In [29] and [30], the authors optimized the joint radio access and power resource allocation to maximize EE while guaranteeing the transmission delay requirements and transmit power constraints of a huge number of devices. To mitigate co-channel interference and further enhance the EE performance of NOMA-based systems  with massive IoT devices, subchannel allocation and power control approaches were proposed in [19], [29], [31]. Furthermore, Miao $et~al.$ [32] proposed an energy-efficient clustering scheme to address spectrum access problem for massive M2M communications. Although the authors in [29]-[32] mainly focused on the EE maximization based massive access, the different QoS requirements (such as latency and reliability) of devices has not been well studied in massive access scenario.

 Considering that intelligence is an important characteristic of future wireless networks, many studies have investigated  application of reinforcement learning (RL) in the field of massive access management recently [9], [23], [33]-[42]. Different distributed RL frameworks were proposed to address the massive access management problem under massive scale and stringent resource constraints [33], [34], where each device has the ability to intelligently make its informed transmission decision by itself without central controller. The authors in [9] and [23] adopted the sparse dictionary learning to facilitate massive connectivity for a massive-device multiple access communication system, and the learning structure does not need any prior knowledge of active devices.  Furthermore, the delay-aware access control of massive random access for mMTC and M2M was studied in [33], [35] and [36], and spectrum access algorithms based on RL were proposed to determine the access decision with high successful connections and low network access latency. As future wireless networks are complex and large-scale, RL cannot effectivity deal with the high-dimensional input state space,  deep reinforcement learning (DRL) (DRL combines deep learning with RL to learn itself from experience) was developed  to solve complex spectrum access decision-making tasks under large-state space [37]-[42]. The authors in [37]-[39] proposed distributed dynamic spectrum access (DSA) approaches based on DRL to search the optimal solution for the DSA problem under the large-state space and local observation information. These distributed learning approaches are capable of encouraging devices to make spectrum access decisions according to their own observations without central controller, and hence they have a great potential for finding efficient solutions for real-time services.  Hua $et~al.$  in [40] presented a network-powered deep distributional Q-network to allocate radio resources for diversified services in 5G networks. Moreover, Yu $et~al.$ in [41] investigated a DRL-based multiple access protocol to learn the optimal spectrum access policy with considering service fairness, and Mohammadi $et~al.$ in [42] employed a deep Q-network (DQN) algorithm for cognitive radio underlay DSA which outperforms the distributed multi-resource allocation. However, the above works [37]-[42] did not investigate how to address the massive access management problem in their presented spectrum access approaches based on DRL, and most of the works did not consider the stringent reliability and latency constraints into the optimization problem.

\subsection{Contributions}

Motivated by the above analysis and observations, in order to address the above mentioned challenges in massive access for 5G and B5G wireless networks, this paper not only studies on how to manage the massive access requests from a huge number of devices, but also takes various QoS requirements (ranging from strict low latency and high reliability to minimum data rate) into consideration. Besides, a novel distributed cooperative learning approach based QoS-aware massive access is presented to optimize the joint subchannel assignment and transmission power control strategy without a centralized controller. The main contributions of the paper are summarized as follows:
\begin{itemize}
\item	We formulate a joint subchannel assignment and transmission power control problem for massive access  considering different practical QoS requirements, and the energy-efficient massive access management problem is modelled as a multi-agent RL problem. Hence, each device has the ability to intelligently make its spectrum access decision according to its own instantaneous observations.
\item	A distributed cooperative subchannel assignment and transmission power control approach based on DRL is proposed for the first time to guarantee both the strict reliability and latency requirements on URLLC services in massive access scenario, where the latency constraint is transformed into a data rate constraint which can make the optimization problem tractable. Specifically, a proper QoS-aware reward function is built to cover both the network EE and devices’ QoS requirements into the learning process.
\item In addition, we apply transfer learning and cooperative learning mechanisms to enable communication links to work cooperatively in a distributed cooperative manner, in order to improve the network performance and transmission success probability based on local observation information. In detail, in transfer learning, if a new device joins the network or applies a new service, or one communication link achieves poor performance (e.g,
low QoS satisfaction level or low convergence speed), then it can directly search the expert agent from the neighbors, and utilizes the transfer learning model from the expert agent instead of building a new learning model. In cooperative learning, devices are encouraged to share their selected actions with their neighbors and take turns to make decisions, which can
enhance the overall benefit by choosing the actions jointly instead of independently.
\item	Extensive simulation results are presented to verify the effectiveness of the proposed distributed cooperative learning approach in massive access scenario, and demonstrate the superiority of the proposed learning approach in terms of meeting the network EE and improving the transmission success probability compared with other existing approaches.
\end{itemize}

The rest of this paper is organized as follows. In Section II, the system model and problem formulation are provided. The massive access management problem is modelled as a Markov decision making process in Section III. Section IV proposes a distributed cooperative multi-agent learning based massive access approach. Section V provides simulation results and Section VI concludes the paper.

\section{System Model and Problem Formulation}

 \vspace{-2pt}
\begin{figure}
\centering
\includegraphics[width=0.55\columnwidth]{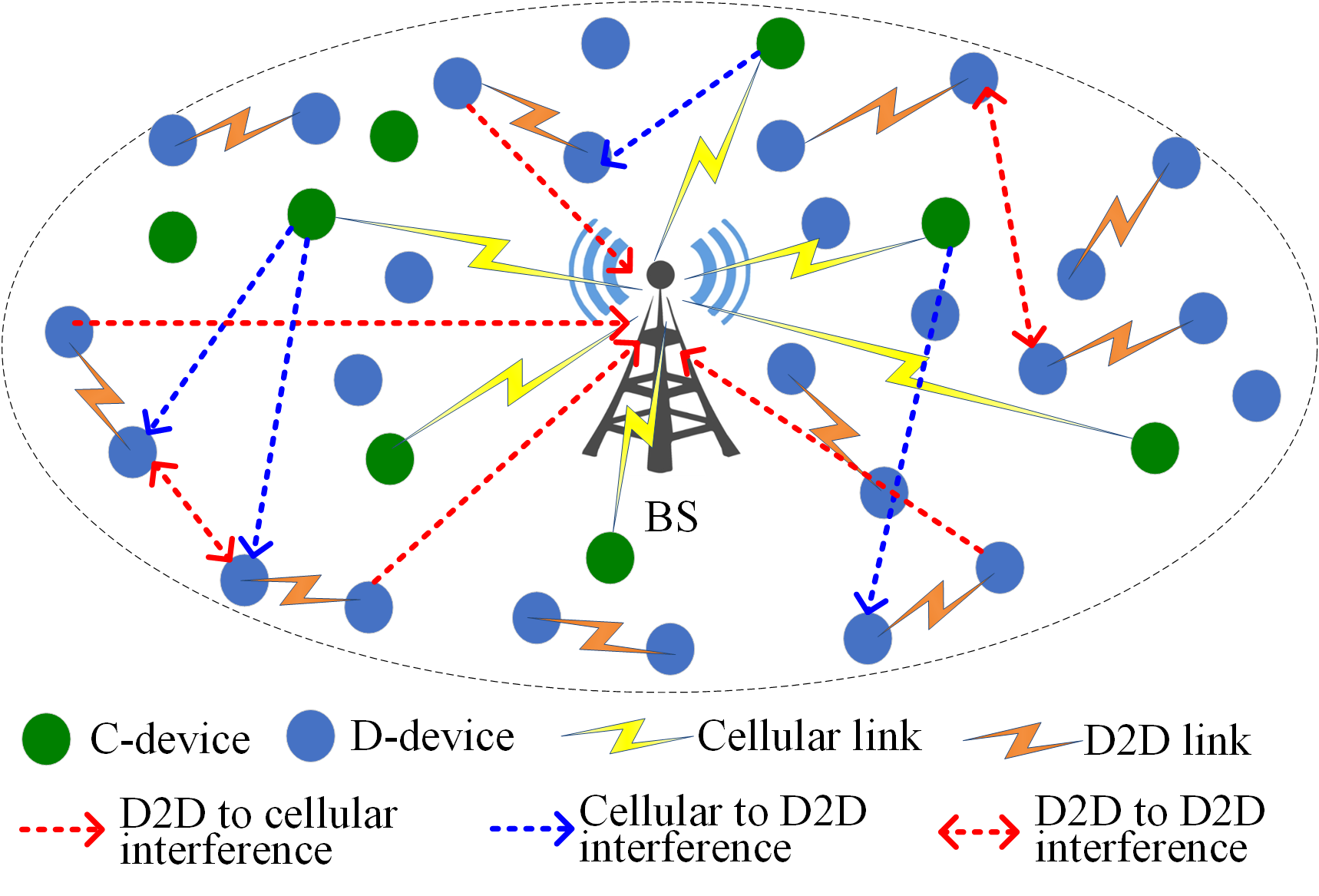}
\vspace{-2pt} \caption{{\small System model of the massive-device network.} } \label{fig:Schematic} \vspace{-5pt}
\end{figure}

We consider a wireless network, as shown in Fig. 1, which
consists of a base station (BS) at the center and a massive number
of devices with each device being equipped with a single
antenna. The devices are mainly divided into two types:
cellular devices (denoted C-device) which communicate with the BS
over the orthogonal spectrum subchannels, and D2D devices
(D-device) which establish D2D communication links if two of them
want to communicate with each other and they are close enough.  In the network,
D-devices can opportunistically access subchannels of
C-devices while ensuring that the generated interference from D2D
pairs to C-devices should not affect the QoS requirements of
C-devices. We assume that each C-device can be allocated with
multiple subchannels, and each subchannel only serve for at most
one C-device in one time slot. In addition, each D2D pair can
share multiple subchannels of C-devices. 

Let $K$, $M$ and $N$ denote the number of C-devices, D2D pairs and
subchannels, respectively. The sets of corresponding C-device , D2D
pair and subchannel  are denoted by ${\mathcal{K}} = \{ 1,2,...,K\} $,
${\mathcal{M}} = \{ 1,2,...,M\} $ and  ${\mathcal{N}} = \{ 1,2,...,N\} $, respectively.
Let $Z$ denote the total number of communication links, $Z$ = $K$
+ $M$, and its corresponding communication link set is defined by
${\mathcal{Z}} = \{ 1,2,...,Z\} $. Denote by ${h_k}$ and ${h_m}$  the channel
coefficients of the desired transmission links from the $k$-th
C-device to the BS, and the transmitter to the receiver in the
$m$-th D2D pair, respectively. Denote by ${g_{k,m}}$, ${g_{m,B}}$
and ${g_{m',m}}$  the interference channel gains from the $k$-th
C-device to the receiver of D2D pair $m$, the transmitter of D2D
pair $m$ to the BS, and the transmitter of the $m'$-th D2D pair to
the receiver of the $m$-th D2D pair, respectively.

 In the spectrum reusing case,
C-devices suffer co-channel interference from the transmitters of
D2D pairs if they share the subchannels with D2D pairs. As a
result, the received signal-to-interference-plus-noise ratio
(SINR) at the BS for C-device $k$ on the $n$-th subchannel is
expressed by
\begin{equation}
\begin{split}
SINR_{k,n}^{\rm{c}} =
\frac{{P_{k,n}^{\rm{c}}{h_k}}}{{\sum\nolimits_{m \in {\mathcal{M}}} {{\rho
_{m,n}}P_{m,n}^{\rm{d}}{g_{m,B}}}  + \delta _k^2}},
\end{split}
\end{equation}
where $P_{m,n}^{\rm{c}}$ and $P_{m,n}^{\rm{d}}$ denote the
transmission power values of the $k$-th C-device and the $m$-th
D2D pair' transmitter on the $n$-th subchannel, respectively.
${\rho _{m,n}}$ is the subchannel access indicator, ${\rho _{m,n}}
\in \{ 0,1\} $; ${\rho _{m,n}} = 1$ indicates that the $m$-th D2D
pair assigns on the $n$-th subchannel; otherwise, ${\rho _{m,n}} =
0$.  $\delta _k^2$ is the additive white Gaussian noise power. In (1), $\sum\nolimits_{m \in {\mathcal{M}}} {{\rho
_{m,n}}P_{m,n}^{\rm{d}}{g_{m,B}}}$ is the co-channel interference.

In addition, subchannel sharing also leads to the co-channel
interference to D2D pairs, which is the generated interference
from the co-channel C-device and co-channel D2D pairs on the same
subchannel. Hence, the received SINR at the $m$-th D2D pair's
receiver when it reuses the $n$-th subchannel of the $k$-th C-device
is given by
\begin{align} 
& SINR_{m,n}^{\rm{d}}\nonumber \\
& = \frac{{P_{m,n}^{\rm{d}}{h_m}}}{{P_{k,n}^{\rm{c}}{g_{k,m}} + \sum\limits_{m' \in \mathcal{M},m' \ne m} {{\rho _{m,m',n}}P_{m',n}^{\rm{d}}{g_{m',m}}}  + \delta _m^2}} ,
\end{align}
where ${\rho _{m,m',n}}$ is the subchannel access indicator,
${\rho _{m,m',n}} \in \{ 0,1\} $ ; ${\rho _{m,m',n}} = 1$
indicates that both the $m$-th D2D pair and  $m'$-th D2D pair assign
on the same $n$-th subchannel in one time slot; otherwise, ${\rho
_{m,m',n}} = 0$. $\delta _m^2$ is the additive white Gaussian
noise power.

Then, the data rate of the $k$-th C-device and the $m$-th D2D pair
on their assigned subchannels are respectively expressed by
\begin{equation}
\begin{split}
R_k^{\rm{c}} = \sum\nolimits_{n \in {\mathcal{N}}} {{\rho _{k,n}}{{\log }_2}(1
+ SINR_{k,n}^{\rm{c}})},
\end{split}
\end{equation} 
and
\begin{equation}
\begin{split}
R_m^{\rm{d}} = \sum\nolimits_{n \in {\mathcal{N}}} {{\rho _{m,n}}{{\log }_2}(1
+ SINR_{m,n}^{\rm{d}})}.
\end{split}
\end{equation}
where ${\rho _{k,n}}$ is the subchannel access indicator which has
the same definition of ${\rho _{m,n}}$  and ${\rho _{m,m',n}}$
as aforementioned above.

\subsection{Network Requirements}
\emph{1) URLLC Requirements:} In 5G and B5G networks, different
devices have different QoS requirements, i.e., some devices have
ultra-high reliability communication requirements, some devices need
strict low-latency services, and even some devices have both the
stringent low latency and high reliability requirements. For
example, intelligent transportation and factory automation
have stringent URLLC requirements for real-time safety information
exchange or hazard monitoring, where the maximum latency is less
than 5 ms (even about 0.1 ms) and the transmission reliability
needs to be higher than that $1-10^{-5}$ (or even $1-10^{-5}$), but they
do not need the high data rate.

For URLLC requirements, we assume that the packet arrival process
of the $i$-th $(i \in {\mathcal{Z}})$ communication link is independent and
identically distributed and follows Poisson
distribution with the arrival rate ${\lambda _i}$ [11], [19]. Let
$L_i^{{\rm{packet}}}$ denote the packet size in bits of the
$i$-th communication link, and it follows the exponential
distribution with mean packet size $\bar L_i^{{\rm{packet}}}$.
Generally, the total latency mainly includes the transmission
delay (${T_{{\rm{tr}}}}$), queuing waiting delay
(${T_{{\rm{qw}}}}$) and processing/computing delay
(${T_{{\rm{pc}}}}$), which can be expressed by [19]
\begin{equation}
\begin{split}
{T_{{\rm{Latency}}}} = {T_{{\rm{tr}}}} + {T_{{\rm{qw}}}} +
{T_{{\rm{pc}}}}.
\end{split}
\end{equation}

In (5), the transmission delay of the packet $L_i^{{\rm{packet}}}$
can be given by  ${T_{{\rm{tr}}}} =
L_i^{{\rm{packet}}}/(W\times{R_i})$, where $W$ is the bandwidth of each
subchannel and ${R_i}$ is the data rate given in (3) or (4),
respectively.

Due to the low latency constraint, each packet requires to be
successfully transmitted in a given time period. Let ${T_{\max }}$ denote the
maximum tolerable latency threshold, so the latency outage
probability of URLLC can be given by
\begin{equation}
\begin{split}
p_i^{{\rm{Latency}}} = \Pr \{ {T_{{\rm{Latency}}}} > {T_{\max }}\}
\le p_{\max }^{{\rm{Latency}}}
\end{split},
\end{equation}
where $p_{\max }^{{\rm{Latency}}}$ is the maximum SINR violation
probability.

It is hard to directly calculate the device's packet latency shown
in (5), and hence the outrage probability in (6) is difficult to
be achieved. However, we can transform the latency constraint (6)
into the data rate constraint by using max-plus queuing
methods [43].

To guarantee the latency outage probability
constraint shown in (6), the data rate $R_i^{{\rm{URLLC}}}$ of each URLLC service of
the $i$-th communication link should meet
\begin{equation}
\begin{split}
R_i^{{\rm{URLLC}}} \ge \frac{{\bar
L_i^{{\rm{packet}}}}}{{W{T_{\max }}}}[{F_i} - {f_{ - 1}}(p_{\max
}^{{\rm{Lat}}}{F_i}{e^{{F_i}}})] \buildrel \Delta \over =
R_{i,\min }^{{\rm{URLLC}}}
\end{split},
\end{equation}
where  ${f_{ - 1}}( \cdot ):[ - {e^{ - 1}},0) \to [ - 1,\infty )]$
denotes the lower branch of Lambert function meeting $y = {f_{ -
1}}(y{e^y})$ [43], ${F_i} = {\lambda _i}{T_{\max }}/(1 -
{e^{{\lambda _i}{T_{\max }}}})$ , and $R_{i,\min }^{{\rm{URLLC}}}$
is the minimum data rate to ensure the latency
constraint shown in (6). The relevant proof of (7) can be seen in [43, Th. 2]. If the transmission data rate is less than the minimum data rate threshold, in other words, the transmission latency exceeds the maximum latency threshold, the current URLLC service is unsuccessful and its corresponding packet transmission is stopped.

In addition, the SINR value can be used to characterize the
reliability of URLLC. In detail, the received SINR at the receiver
should be beyond the minimum SINR threshold. Otherwise, the received
signal cannot be successfully demodulated. Hence, the outage
probability in term of SINR can be given by
\begin{equation}
\begin{split}
p_{i,n}^{{\rm{outage}}} = \Pr \{ SIN{R_{i,n}} < SINR_{i,n}^{\min
}\}  \le p_{\max }^{{\rm{outage}}}
\end{split},
\end{equation}
where $SINR_{i,n}^{\min }$ denotes the minimum SINR threshold of
communication link $i$ on the $n$-th subchannel and
$p_{\max }^{{\rm{outage}}}$ denotes the maximum violation
probability.

\emph{2) Minimum Data Rate Requirements:} In addition to the high reliability and low latency requirements
mentioned in Section, some C-devices and D2D pairs may have the
minimum data rate requirements. Let  $R_{k,\min }^{\rm{c}}$ and
$R_{m,\min }^{\rm{d}}$ denote the minimum data rate requirements
of the $k$-th C-device and the $m$-th D2D pair, respectively. Then,
the minimum data rate requirements are given by
\begin{equation}
\begin{split}
R_k^{\rm{c}} \ge R_{k,\min }^{\rm{c}},\;\forall k;\;\;
R_m^{\rm{d}} \ge R_{m,\min }^{\rm{d}},\;\forall m.
\end{split}
\end{equation}

\subsection{Problem Formulation}

The objective of this paper is to maximize the overall network EE
(EE is the ratio of the sum data rate and the sum energy
consumption) while guaranteeing the network requirements shown in
Section III.A. Then, the massive access management problem (joint
subchannel access and transmission power control) is formulated as follows:
\begin{equation}
\begin{split}
\begin{array}{l}
\mathop {\max }\limits_{{\bm{\rho }},{\bm{P}}} \;\;\;{\eta _{EE}} = \frac{{\sum\nolimits_{k \in {\mathcal{K}}} {R_k^{\rm{c}}}  + \sum\nolimits_{m \in {\mathcal{M}}} {R_m^{\rm{d}}} }}{{\sum\limits_{n \in {\mathcal{N}}} {\left( {\sum\limits_{k \in {\mathcal{K}}} {{\rho _{k,n}}P_{k,n}^{\rm{c}}}  + \sum\limits_{m \in {\mathcal{M}}} {{\rho _{m,n}}P_{m,n}^{\rm{d}}} } \right)}  + Z{P_{cir}}}}\\
s.t.\;\;({\rm{a}}):\;(7),\;(8),\;(9);\;\\
\;\;\;\;\;\;\;({\rm{b}}):\;{\rho _{k,n}} \in \{ 0,1\} ,\;\;{\rho _{m,n}} \in \{ 0,1\} ,\;\forall k,\;m,\;n;\\
\;\;\;\;\;\;\;({\rm{c}}):\;\sum\nolimits_{k \in {\mathcal{K}}} {{\rho _{n,k}}}  \le 1,\,\,\forall n \in {\mathcal{N}};\;\\
\;\;\;\;\;\;\;({\rm{d}}):\;\sum\nolimits_{n \in {\mathcal{N}}} {{\rho _{k,n}}P_{k,n}^{\rm{c}}}  \le P_{\max }^{\rm{c}},\;\forall k \in {\mathcal{K}};\\
\;\;\;\;\;\;\;({\rm{e}}):\;\sum\nolimits_{n \in {\mathcal{N}}} {{\rho
_{m,n}}P_{m,n}^{\rm{d}}}  \le P_{\max }^{\rm{d}},\;\forall m \in
{\mathcal{M}},
\end{array}
\end{split}
\end{equation}
where ${\bm{\rho }}$ and ${\bm{P}}$ denote the subchannel
assignment and power control strategies, respectively. $P_{\max
}^{\rm{c}}$ and $P_{\max }^{\rm{d}}$ denote the maximum transmission
power values of each C-device and each D-device, respectively.
${P_{cir}}$ denotes the circuit power consumption of one
communication link. Constraint (10c) guarantees that each
subchannel is allocated at most one C-device. Constraint (10d) and
(10e) are imposed to ensure that the power constraints of devices.

\section{Problem Transformation}

Clearly, the optimization problem given  in (10) is not easy to be
solved as it is a non-convex combination and NP-hard problem.
More importantly, the optimization objective is just a single time
slot optimization problem, where the massive access decision is
only based on the current state with the fixed optimization
function. The single time slot massive access decision approaches
may converge to the suboptimal solution and obtain the
greedy-search like performance due to the lack of the
historical network state and the long term benefit. 

Hence, in this section, model-free RL as a dynamic programming
tool  can be applied to address the decision-making problem
by learning the optimal solutions over dynamic environment.
Similar to most of existing studies [32]-[42], we apply Markov
Decision Process (MDP) to model the massive access decision-making
problem in the RL framework by transforming the optimization
problem (10) into MDP. 

In the MDP model, each communication link acts as an agent by
interacting with outside environment and the MDP model is defined
as a tuple  $({\mathcal{S}},{\mathcal{A}},{\mathcal{P}},r,\gamma )$, where ${\mathcal{S}}$ is the state
space set,   ${\mathcal{A}}$ denotes the action space set,  ${\mathcal{P}}$ indicates the
transition probability: ${\mathcal{P}}({s_{t + 1}}|{s_t},{a_t})$ is the
probability of transferring from a current state ${s_t} \in {\mathcal{S}}$ to
a new state ${s_{t + 1}} \in {\mathcal{S}}$ after taking an action ${a_t} \in
{\mathcal{A}}$, $r$ denotes the immediate reward, and $\gamma  \in (0,1)$ denotes the
discount factor. The details of the MDP model for massive access
management are presented as follows.

\textbf{State:} In 5G and B5G networks, the network state is
defined as  $s = \{
{s_{{\rm{cha}}}},{s_{{\rm{cq}}}},{s_{{\rm{tr}}}},{s_{{\rm{QoS}}}}\}
\in {\mathcal{S}}$, ${s_{{\rm{cha}}}}$ indicates the subchannel working status
(idle or busy); ${s_{{\rm{cq}}}}$ depicts the channel quality
(i.e., SINR); ${s_{{\rm{tr}}}}$ is the traffic load of each
packet; and ${s_{{\rm{QoS}}}}$ represents the QoS satisfaction
level (the transmission success probability), such as the
satisfaction levels of the minimum data rate, latency and
reliability.

\textbf{Action:} For the massive access management problem, each
agent will decide which subchannels can be assigned and how
much transmit power should be allocated on the assigned
subchannels. Hence, the action can be defined as $a = \{ {\rho
_{{\rm{cha}}}},{P_{{\rm{pow}}}}\}  \in {\mathcal{A}}$ which includes the
subchannel assignment indicator (${\rho _{{\rm{cha}}}}$) and the
transmission power (${P_{{\rm{pow}}}}$). At each time slot, the action of each device consists of channel assignment indicator $ {\rho _{{\rm{cha}}}} \in \{0,1\}$ and transmission power level $ {P_{{\rm{pow}}}} \in \{50, 150, 300, 500\}$ in mW where the transmission power is discretized into four levels. We can observe that the action space of each device is not big, but the overall action space of all devices in the massive access scenario is large. Hence, we need to discretize the transmission power numbers as small as possible, so the four transmission power levels are chosen in this paper instead of the higher number of transmission power levels.

\textbf{Reward function:} In order to reflect the device
experience which the network wants to optimize, RL requires
designing the specific reward function where the learning process
is generally driven by the reward. In RL, each agent searches its
decision-making policy by maximizing its reward under the
interaction with environment. Hence, it is important to design an
efficient reward function to improve the devices' service
satisfaction levels.

Here, let ${\mathcal{Z}}'$ denotes the set of communication links in the URLLC
scenario where the devices have both the reliability and latency
requirements, and ${\mathcal{Z}}''$  denotes the set of communication links in
the normal scenario where the devices have minimum data
requirements. $|{\mathcal{Z}}'| = Z'$ and $|{\mathcal{Z}}''| = Z''$. Let
$R_i^{{\rm{nor}}}$ and $R_{i,\min }^{{\rm{nor}}}$ denote the
instantaneous data rate and the minimum data rate threshold in the
normal scenario, respectively.

According the optimization problem shown in (10), considering the
different QoS requirements, we design a new QoS-aware reward
function for the massive access management problem, where the
reward function of the $i$-th communication link includes the network EE, as well as the
reliability, latency and minimum data rate requirements, which is
expressed by
\begin{equation}
\begin{split}
r_{i} =  {{\eta _{i, EE}}} -
 {{c_1} {\chi
_i^{{\rm{URLLC}}}} }- 
{{c_2} {\chi
_i^{{\rm{nor}}}} },
\end{split}
\end{equation} 
where
\begin{equation}
\begin{split}
\chi _i^{{\rm{URLLC}}} = \left\{ \begin{array}{l}
1,\;\;{\rm{if}}\;\;(7)\;{\rm{or}}\;(8)\;{\rm{is}}\;{\rm{not}}\;{\rm{satisfied}},\;\\
0,\;{\rm{otherwise}}{\rm{,}}
\end{array} \right.
\end{split}
\end{equation}
and
\begin{equation}
\begin{split}
\chi _i^{{\rm{nor}}} = \left\{ \begin{array}{l}
1,\;\;{\rm{if}}\;R_i^{{\rm{nor}}} < R_{i,\min }^{{\rm{nor}}}\;,\;\\
0,\;{\rm{otherwise}}{\rm{.}}
\end{array} \right.
\end{split}
\end{equation}

In (11), the part 1 indicates the immediate utility (network EE),
the part 2 and part 3 are the cost functions of the transmission failures which are defined as the unsatisfied URLLC requirements and the unsatisfied minimum data
rate requirements, respectively.  The
parameters ${c_i}$, $i \in \{ 1,2\} $ denote the positive constants
of the latter two parts in (11) and they are adopted for balancing
the utility and cost [19], [28], [39].

The objectives of (12) and (13) are to refract the QoS
satisfaction levels of both the URLLC services and normal
services, respectively. In detail, if the URLLC requirement of one packet is
satisfied in the current time slot, then $\chi _i^{{\rm{URLLC}}} =
0$; if the minimum data rate is satisfied, then $\chi _i^{{\rm{nor}}} = 0$. This means that there is no
cost or punishment of the reward due to the successful
transmission with QoS guarantees. Otherwise, $\chi _i^{{\rm{URLLC}}} =
1$, or $\chi _i^{{\rm{nor}}} = 1$.
  
The reward function shown in (11) may have the same reward values for some cases. For example, the following two cases may have the same reward for different values: Case I, the URLLC requirement is not satisfied while the minimum data rate requirement is satisfied, then  ${\chi ^{{\rm{URLLC}}}} = 1$ and  ${\chi ^{{\rm{nor}}}} = 0$; Case II, the URLLC requirement is satisfied while the minimum data rate requirement is not satisfied, then  ${\chi ^{{\rm{URLLC}}}} = 0$ and  ${\chi ^{{\rm{nor}}}} = 1$. For these two cases, they may have the same reward function values:  $r = {\eta _{EE}} - {c_1} * 1 - {c_2} * 0$ and $r = {\eta _{EE}} - {c_1} * 0 - {c_2} * 1$ with ${c_1} = {c_2}$  being the punishment factors.  If the punishment factors  ${c_1} \ne {c_2}$,  the two cases have different reward function values. We would like to mention that the values of the punishment factors ${c_1}$  and ${c_2}$ have important impacts on the reward function, if  ${c_1} > {c_2}$, the  URLLC requirement has the higher impact on the final reward value than that of the minimum data rate requirement; by contract,  if ${c_1} < {c_2}$, the minimum data rate requirement has the higher impact on the final reward value than that of the URLLC requirement. Furthermore, if  ${c_1} = {c_2}$, both the URLLC requirement and minimum data rate requirement have the same impacts on the reward value.


In RL, each agent in the MPD model tries to select a policy $\pi $
to maximize a discounted accumulative reward, where $\pi $  is a
mapping from state $s$ with the probability distribution over
actions that the agent can take: $\pi (s):{\mathcal{S}} \to {\mathcal{A}}$. The discounted
accumulative reward is also a called the state-value function for
starting the state $s$ with the current policy $\pi $, and it is  defined by
\begin{equation}
\begin{split}
{V^\pi }(s) = \left\{ {\sum\limits_{t = 1}^\infty  {{\gamma
^t}{r_t}({s_t},{a_t})|} {s_0} = s,\pi } \right\}.
\end{split}
\end{equation}

The function ${V^\pi }(s)$  in (14) is usually applied to test the
quality of the selected policy $\pi $  when the agent selects the
action $a$. The MPD model tries to search the
optimal state-value function ${V^ * }(s)$, which is expressed by
\begin{equation}
\begin{split}
{V^ * }(s) = \mathop {\max }\limits_\pi  {V^\pi }(s).
\end{split}
\end{equation}

Once ${V^ * }(s)$ is achieved, the optimal policy  ${\pi ^ * }({s_t})$ under the current state ${s_t}$ is
determined by
\begin{equation}
\begin{split}
{\pi ^ * }({s_t}) = \arg \mathop {\max }\limits_{{a_t} \in {\mathcal{A}}
}
{\bar U_t}({s_t},{a_t}) + \sum\limits_{{s_{t + 1}}} {P({s_{t +
1}}|{s_t},{a_t})} {V^ * }({s_{t + 1}}),
\end{split}
\end{equation}
where ${\bar U_t}({s_t},{a_t})$ denotes the expected reward by
selecting action ${a_t}$ at state ${s_t}$. To calculate  ${V^
* }(s)$, the iterative  algorithms can be applied. However, it is
difficult to get the transition probability  $P({s_{t +
1}}|{s_t},{a_t})$ in practical environments, but RL algorithms,
such as Q-learning, policy gradient and DQN, are widely
employed to address MDP problems under environment uncertainty.

In Q-learning algorithm, the Q-function is used to calculate the accumulative reward for starting from a state $s$ by taking
an action $a$ with the selected policy $\pi $, which can be given
by
\begin{equation}
\begin{split}
{Q^\pi }(s,a) = \left\{ {\sum\limits_{t = 1}^\infty  {{\gamma
^t}{r_t}({s_t},{a_t})|} {s_0} = s,{a_0} = a,\pi } \right\}.
\end{split}
\end{equation}

Similarly, the optimal Q-function is obtained by
\begin{equation}
\begin{split}
{Q^ * }(s,a) = \mathop {\max }\limits_\pi  {V^\pi }(s,a).
\end{split}
\end{equation}

In Q-learning algorithm, the Q-function is updated by
\begin{equation}
\begin{split}
\begin{array}{l}
{Q_{t + 1}}({s_t},{a_t}) = {Q_t}({s_t},{a_t})\\
 + \alpha \left[ {{r_{t + 1}} + \gamma \mathop {\max }\limits_{{a_{t + 1}}} {Q_t}({s_{t + 1}},{a_{t + 1}}) - {Q_t}({s_t},{a_t})} \right],
\end{array}
\end{split}
\end{equation}
where $\alpha $ denotes the learning rate. When ${Q^ * }(s,a)$  is achieved, the optimal policy is
determine by
\begin{equation}
\begin{split}
{\pi ^ * }(s) = \arg \mathop {\max }\limits_{a \in A} {Q^ * }(s,a).
\end{split}
\end{equation}

\section{Distributed Cooperative
Multi-Agent RL Based Massive Access}

Even through Q-learning is widely adopted to design the resource
management policy in wireless networks without knowing the
transition probability in advance, it has some key limitations for
its application in large-scale 5G and B5G networks, such as
Q-learning has slow convergence speed under large-state space, and
it cannot deal with large continuous state-action spaces.
Recently, a great potential is demonstrated by DRL that combines
neural networks (NNs) with Q-learning, called DQN, which
can efficiently address the above mentioned problems and achieve
better performance owing to the following reasons. Firstly, DQN
adopts NNs to map from the observed state to action between
different layers, instead of using storage memory to store the Q-values.
Secondly, large-scale models can be represented from high
dimensional raw data by using NNs. Furthermore, applying
experience replay and generalization capability brought by NNs,
DQN can improve network performance.

In 5G and B5G networks shown in Fig. 1, massive communication
links aim to access the limited radio spectrum, which can be
modelled as a multi-agent RL problem, where each communication link
is regarded as a learning agent to interact with
network environment to learn its experience, and the learned
experience is then utilized to optimize its own spectrum access
strategy. Massive agents explore the outside network environment
and search spectrum access and power control strategies according
to the observations of the network state.

The proposed deep multi-agent RL based approach consists of two
stages, a training stage and a distributed cooperative
implementation stage. The main contributions of the proposed distributed cooperative multi-agent RL based approach for massive access are provided in the following two subsections in detail.

\subsection{Training Stage of Multi-Agent RL for Massive Access}

For the training sage, we adopt DQN with experience relay to train
the multi-agent RL for efficient learning of massive access
policies. Fig. 2 indicates the training process. All communication
links are regarded as agents and the wireless network acts as the
environment. Firstly, each agent intelligently observes its
current state (e.g., subchannel status (busy or idle), channel
quality, traffic load and QoS satisfaction levels) by integrating
with the environment. Then, it makes decision and chooses one
action according to its learned policy. After that, the
environment feedbacks a new state and an immediate reward to each
agent. Based on the feedback, all agents smartly learn new
policies in the next time step. The optimal parameters of DQN can
be trained with an infinite number of time steps. In addition, the
experience replay mechanism is adopted to improve the learning
speed, the learning efficiency and the learning stability
toward the optimal policy for the massive access management. The
training data is stored in the storage memory, and a random
mini-batch data is sampled from the storage memory and used to
optimize the weight of DQN.

 \vspace{-2pt}
\begin{figure}
\centering
\includegraphics[width=0.8\columnwidth]{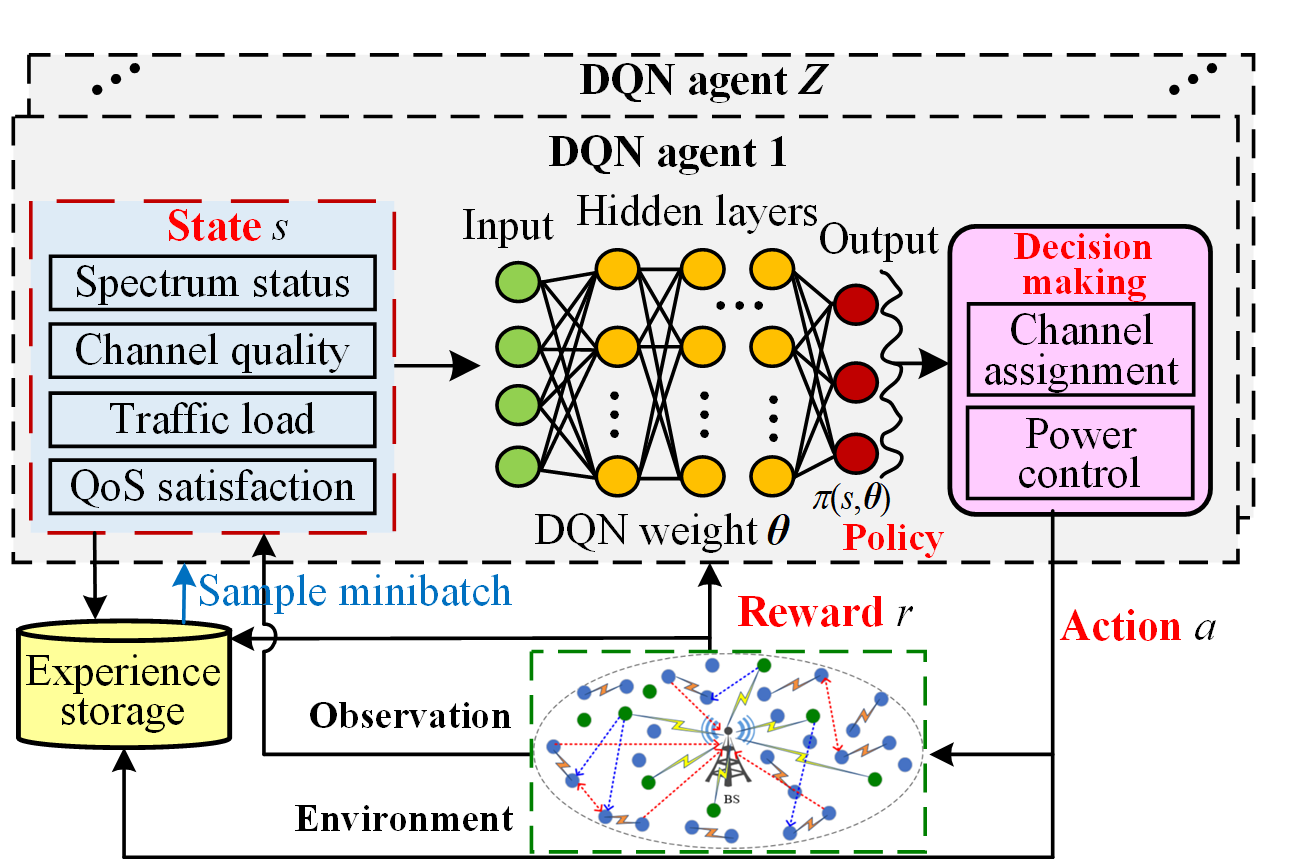}
\vspace{-2pt} \caption{{\small DQN training based intelligent
subchannel assignment and power control for massive access.} }
\label{fig:Schematic} \vspace{-5pt}
\end{figure}

At each training or learning step, each DQN agent updates its
weight, $\bm{\theta} $ , to minimize the loss function defined by
\begin{equation}
\begin{split}
\begin{array}{l}
Loss({{\bm{\theta} _t}}) = \\
{\left[ {{r_{t + 1}}({s_t},{a_t}) + \gamma \mathop {\max }\limits_{a \in \mathcal{A}} {Q_t}({s_{t + 1}},{a_{t + 1}},{{\bm{\theta} _t}}) - {Q_t}({s_t},{a_t},{{\bm{\theta} _t}})} \right]^2}.
\end{array}
\end{split}
\end{equation}

One important reason of adopting DQN is to update the loss
functions given in (21) at each tainting step to decrease the
computational complexity for large-scale learning problems [37]-[42].
The DQN weight $\bm{\theta} $ is obtained by using the gradient descent
method, which can be expressed as
\begin{equation}
\begin{split}
{\bm{\theta} _{t + 1}} = {\bm{\theta} _t} + \beta \nabla Loss({\bm{\theta}
_t}),
\end{split}
\end{equation}
where $\beta $ denotes the learning rate of the weight $\bm{\theta} $,
and $\nabla (.)$ is the first-order partial derivative.

Then, each agent selects its action according to the selected
policy $\pi {\rm{(}}{s_t},{\bm{\theta} _t}{\rm{)}}$, which is given by
\begin{equation}
\begin{split}
\pi {\rm{(}}{s_t},{\bm{\theta} _t}{\rm{)}} = \arg \mathop {\max
}\limits_{a \in {\mathcal{A}}} \left\{ {{Q_t}({s_t},{a_t},{\bm{\theta} _t})}
\right\}.
\end{split}
\end{equation}

Pseudocode for training DQN is presented in \textbf{Algorithm 1}.
The communication environment contains both the C-devices and
D-devices and their positions in the served coverage area of the
BS, and the channel gains are generated based on their positions.
Each agent has its trained DQN model that takes as input of
current observed state ${s_t}$ and outputs the Q-function
with the selected action  ${a_t}$.  The training loop
has a finite number of episodes  ${N^{{\rm{epi}}}}$ (i.e., tasks)
and each episode has  $T$ training iterations. At each training
step, after observing the current state ${s_t}$, all agents
explore the state-action space by applying the $\varepsilon  -
$greedy method, where each action ${a_t}$ is randomly selected
with the probability ${\varepsilon _t}$ while the action is chosen with the largest
Q-value ${Q_t}({s_t},{a_t},{\bm{\theta} _t})$ with the probability $1 -
{\varepsilon _t}$. After executing ${a_t}$ (subchannel assignment
and power control), agents will receive an immediate reward
${r_t}$ and observe a new state  ${s_{t + 1}}$ from the
environment. Then, the experience ${e_t} =
({s_t},{a_t},r({s_t},{a_t}),{s_{t + 1}})$ is stored into the
replay memory $D$. At each episode, a mini-batch data from the
memory is sampled to update the weight ${\bm{\theta} _t}$ of DQN.

\begin{algorithm}[t]
\begin{small}
\caption{\small DQN Training Stage of Subchannel Assignment and
Power Control with Multi-Agent RL for Massive Access}

1: \textbf{Input:} DQN structure, environment simulator and QoS requirements of all devices (e.g., reliability, latency and minimum data rate). \\
2: \textbf{for}   each episode $j$=1,2,..., ${N^{{\rm{epi}}}}$ \textbf{do}   \\
3: $~$ \textbf{Initialize:} Initial Q-networks for all agents (e.g., Q-function $Q(s,a)$, policy strategy $\pi (s,a)$, and weight $\bm{\theta }$) and  experience replay $D$.\\
4: $~~$ \textbf{for} each iteration step $t$=0,1,2,..., $T$ \textbf{do} \\
5: $~~~$ Each agent observes its state  ${s_t}$;\\
6: $~~~$ Select a random action ${a_t}$ with the probability $\varepsilon $;\\
7: $~~~$ Otherwise, choose the action ${a_t} = \arg \mathop {\max }\limits_{a \in {\mathcal{A}}} {Q_t}({s_t},{a_t},{\bm{\theta _t}})$;\\
8: $~~~$ Execute action ${a_t}$, then obtain a reward  ${r_t}$ by (15), and observe a new state ${s_{t + 1}}$; \\
9: $~~~$ Save experience  ${e_t} = ({s_t},{a_t},r({s_t},{a_t}),{s_{t + 1}})$ into the storage memory $D$; \\
10: $~$ \textbf{end for}\\
11: $~$ \textbf{for} each agent \textbf{do}\\
12: $~~$  Sample a random mini-batch data ${e_t}$  from $D$;\\
13: $~~$  Update the loss function by (21);\\
14: $~~$ Perform a gradient descent step to update ${\bm{\theta} _{t + 1}}$  by (22);\\
15: $~~$ Update the policy  $\pi $ with maximum Q-value by
(23), and chose an action based on $\pi $; \\
16: $~$ \textbf{end for}\\
17:  \textbf{end for}\\
18: \textbf{return:} Return trained DQN models. \\
\end{small}
\label{alg_lirnn}
\end{algorithm}


\subsection{Distributed Cooperative Implementation of Multi-Agent RL for
Massive Access}

 \vspace{-2pt}
\begin{figure}
\centering
\includegraphics[width=0.95\columnwidth]{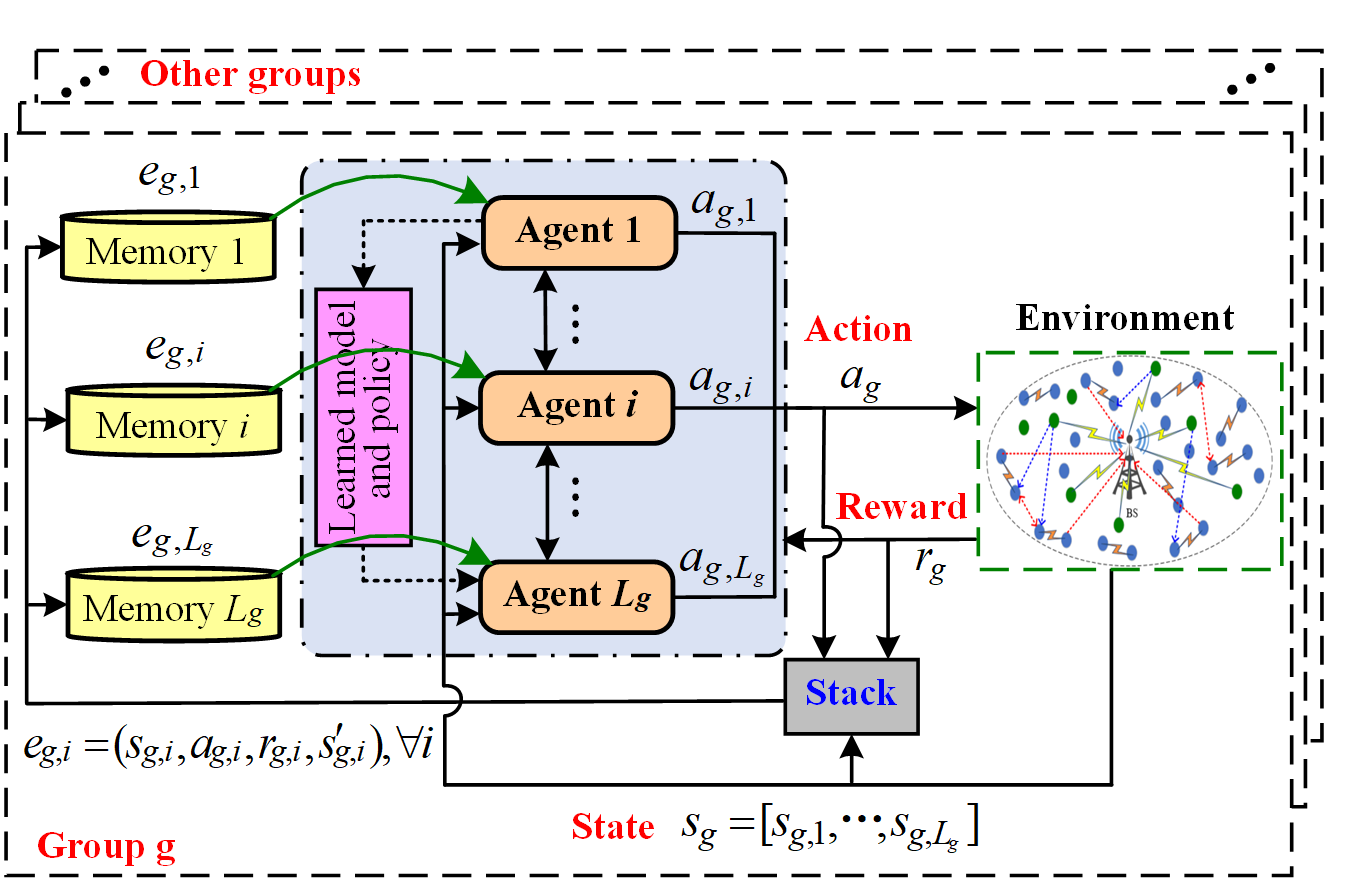}
\vspace{-2pt} \caption{{\small Distributed cooperative multi-agent
RL framework.} } \label{fig:Schematic} \vspace{-5pt}
\end{figure}

The above mentioned trained DQN models with the computation
intensive training procedure are shown in Section IV.A, which can be
completed offline at BS since BS has powerful computing capacity
to train large-scale models. After adequate training, the trained
models are utilized for implementation. In this subsection, we
propose a distributed cooperative learning approach to optimize
the network performance in massive access scenario.

During the distributed cooperative implementation stage, at each
learning step, each communication link (agent) utilizes its local
observation and information to choose its action with the maximum
Q-value. In this case, each agent has no knowledge of actions
chosen by other agents if the actions are updated simultaneously
and new joined agents need to train their own learning model with
extra training computational time or cost.  In order to address
this issue, motivated by the concept of transfer learning and
cooperative learning, we present a distributed cooperative
learning approach to improve the learning efficiency and enhance
the service performance of each agent, where devices are
encouraged to communicate and share their learned experiences and
decisions within a small number of neighbors, and finally learn with
each other, as shown in Fig.3.

\emph{1) Transfer Learning:}

\textbf{(i) The Expert Agent Selection:} When a new device joins
5G and B5G networks, or one device applies a new communication
service, instead of building a new learning model, it can
communicate with neighboring devices to search one suitable expert
to utilize the expert' current learning model. In addition, if one
communication link has poor performance (e.g., low convergence
speed and poor QoS satisfaction levels) according to its current
leaning strategy, it can search one neighboring communication link
(agent) as the expert and then utilizes the learned model or policy from the expert.

Generally, to find the expert, devices exchange the following
several metrics with their neighbors: \emph{a)} the types of
device, e.g., C-device and D2D device; \emph{b)} the communication
services, which mainly refer to URLLC service and normal service;
\emph{c)} the related QoS parameters, such as the target
thresholds of reliability and latency, and the minimum data rate.
The similarity of the agents can be evaluated by adopting the
manifold learning, which is also called Bregman Ball [19]. The
Bregman Ball is defined as the minimum manifold with a central
${\Theta _{{\rm{cen}}}}$ (the information of the learning agent, where information refers to the types of device,communication services, and QoS parameters mentioned above),
and a radius ${\Psi _{{\rm{rad}}}}$. Any information point
${\Theta _{{\rm{poi}}}}$ (the information of neighbors) is inside
this ball, and the agent tries to search the information point which has
the highest similarity with ${\Theta _{{\rm{cen}}}}$. The
distance between any information point and the central ${\Theta
_{{\rm{cen}}}}$ is defined by
\begin{equation}
\begin{split}
{\rm{Dis}}({\Theta _{{\rm{cen}}}},{\Psi _{{\rm{rad}}}}) = \left\{
{{\Theta _{{\rm{poi}}}} \in \Theta :{\rm{Dis}}({\Theta
_{{\rm{poi}}}},{\Theta _{{\rm{cen}}}}) \le {\Psi _{{\rm{rad}}}}}
\right\}.
\end{split}
\end{equation}

After the highest similarity level (the smallest distance achieved by (24)) between the learning agent and
the expert agent is found, the learning agent can use the
learned DQN model of the selected expert agent.

\textbf{(ii) Learning from Expert Agent:} As analyzed above, after
finding the expert agent, the learning agent uses the transferred
DQN model ${Q^{{\rm{Transfer}}}}(s,a)$ from the expert agent and
its current native DQN model ${Q^{{\rm{Current}}}}(s,a)$ to
generate an overall DQN model. Accordingly, the new Q-table of the
learning agent can be expressed as
\begin{equation}
\begin{split}
{Q^{{\rm{New}}}}(s,a) = \mu {Q^{{\rm{Transfer}}}}(s,a) + (1 - \mu
){Q^{{\rm{Current}}}}(s,a),
\end{split}
\end{equation}
where $\mu  \in [0,1]$ is the transfer rate, and it will be
gradually decreased after each learning step to reduce the effect
of the transferred DQN model from the expert agent on the new DQN
model.

In the distributed cooperative manner, the policy vector of all
agents are updated as follows:
\begin{equation}
\begin{split}
{{\bm{\pi }}_{t + 1}}{\rm{(}}{s_t}{\rm{)}} = \left[
\begin{array}{l}
\pi _{t + 1}^1{\rm{(}}s_t^1{\rm{)}}\\
\;\;\;\;\;\;\; \vdots \\
\pi _{t + 1}^i{\rm{(}}s_t^i{\rm{)}}\\
\;\;\;\;\;\;\; \vdots \\
\pi _{t + 1}^Z{\rm{(}}s_t^Z{\rm{)}}
\end{array} \right] = \left[ \begin{array}{l}
\arg \mathop {\max }\limits_{{a^1} \in {{\mathcal{A}}^1}} \left\{ {Q_{t + 1}^1(s_t^1,a_t^1)} \right\}\\
\;\;\;\;\;\;\;\;\;\;\;\;\;\;\;\;\;\;\; \vdots \\
\arg \mathop {\max }\limits_{{a^i} \in {{\mathcal{A}}^i}} \left\{ {Q_{t + 1}^i(s_t^i,a_t^i)} \right\}\\
\;\;\;\;\;\;\;\;\;\;\;\;\;\;\;\;\;\;\; \vdots \\
\arg \mathop {\max }\limits_{{a^Z} \in {{\mathcal{A}}^Z}} \left\{ {Q_{t +
1}^Z(s_t^Z,a_t^Z)} \right\}
\end{array} \right]
\end{split}
\end{equation}
where $Q_{t + 1}^i(s_t^i,a_t^i,\bm{\theta} _t^i)$ denotes the
Q-function of the $i$-th agent (communication link) with its
current state-action pair $(s_t^i,a_t^i)$ at the current time slot
in its DQN model.

When the state-action pairs are  visited for many enough times for convergence, all Q-tables will
converge to the final point  ${Q^ * }$. Hence, we can get the final learned
policy as follow
\begin{equation}
\begin{split}
{{\bm{\pi }}^ * }{\rm{(}}s{\rm{)}} = \left[ \begin{array}{l}
\arg \mathop {\max }\limits_{{a^1} \in {{\mathcal{A}}^1}} \left\{ {{Q^1}^ * ({s^1},{a^1})} \right\}\\
\;\;\;\;\;\;\;\;\;\;\;\;\;\;\;\;\;\;\; \vdots \\
\arg \mathop {\max }\limits_{{a^Z} \in {{\mathcal{A}}^Z}} \left\{ {{Q^Z}^ *
({s^Z},{a^Z},)} \right\}
\end{array} \right].
\end{split}
\end{equation}

\emph{2) Cooperative Learning}

 If the action is chosen independently
according to the local information, each communication link has no
information of actions selected by other communication links when
the actions are updated simultaneously. Consequently, the states
observed by each communication link may fail to fully characterize
the environment. Hence, cooperation and decision sharing among
agents in the proposed distributed learning approach can improve
the network performance, where a small number of communication
links will share their actions with their neighbors. In the
cooperative manner, the massive number of agents can be classified
into $G$ groups, where the $g$-th group consists of ${L_g}$ agents
and the agents in the same group are also their neighboring
agents. The group division principle can adopt the studies in [13], [26] and [32].

In general, it is possible to approximate the sum utility of the
$g$-th group  ${Q_g}({s_g},{a_g})$ by the sum of each agent'
utility ${Q_{g,i}}({s_{g,i}},{a_{g,i}})$ in the same group, where
$s_g$ and $a_g$ denote the entire state and action of the $g$-th group,
respectively;  ${s_{g,i}}$ and ${a_{g,i}}$ are the individual
state and action of the $i$-th agent in the $g$-th group,
respectively. Hence, the total utility in a small group $g$ can be
calculated by
\begin{equation}
\begin{split}
{Q_g}({s_g},{a_g}) = \sum\nolimits_{i = 1}^{{L_g}}
{{Q_{g,i}}({s_{g,i}},{a_{g,i}})}.
\end{split}
\end{equation}

  Then, the joint optimal policy learned in the $g$-th group can be expressed by
\begin{equation}
\begin{split}
{\pi _g}{\rm{(}}{s_g}{\rm{)}} = \arg \mathop {\max }\limits_{{a_g}
\in {{\mathcal{A}}_g}} \left\{ {{Q_g}({s_g},{a_g})} \right\},
\end{split}
\end{equation}
where ${{\mathcal{A}}_g}$ denotes the entire action space of the $g$-th group.

In fact, the cooperation can be defined by allowing communication
links (agents) to share their selected actions with their
neighboring links and take turns to make decisions, which can
enhance the overall feedback reward by choosing the actions
jointly instead of independently. For example, in the fully
distributed learning manner, each spectrum access may run into
collisions when other links make their decisions independently and
happen to assign the same subchannel, leading to the increased
co-channel interference and reduce the performance. By
contrast, in the cooperative learning scenario, to avoid such
situation, each communication link has information of the
neighbors' actions in its observation, and try to avoid the assignment of
the same subchannel in order to achieve more rewards.

The distributed cooperative implementation of multi-agent RL for
massive access is shown in \textbf{Algorithm 2}. Generally, at
each time step, after observing the states (subchannel
occupation status, channel quality, traffic load, QoS satisfaction
level, etc.) from the environment,  the actions (massive subchannel assignment and
power control) in communication links are selected with the
maximum Q-value given by loading the trained DQN models shown in
\textbf{Algorithm 1}. As mentioned above, a small number of
neighboring devices are encouraged to cooperate with each other in
the same group to maximize the sum Q-value shown in (28), where
their decisions are shared in the same group and the joint action
strategy ${a_g}$ is selected with the maximum cooperative Q-value.
In addition, it is worth noting that if a new device joins
the network or applies a new service, or one communication link
achieves poor performance (e.g, low transmission success
probability or low convergence speed), then it can directly search
the expert agent from the neighbors in the same group, and
utilizes the transfer learning model and policy from the expert
agent. Finally, all communication links begin transmission with
the subchannel assignment and transmission power strategies
determined by their learned policies.

\begin{algorithm}[t]
\begin{small}
\caption{\small Distributed Cooperative Implementation of
Multi-Agent RL for Massive Access}

1: \textbf{Input:} DQN structure, environment simulator and QoS requirements of all devices. \\
2: \textbf{start:}   Load DQN models.  \\
3: \textbf{loop}\\
4: $~$ Each agent (communication link) observes its state $s$; \\
$~~~~$ \textbf{\emph{Transfer learning}}\\
5: $~$ \textbf{if} the agent is new, or needs new service or has poor performance, \textbf{then}; \\
6: $~~~$ The agent exchanges information with its neighbors;\\
7: $~~~$ Search the expert with the highest similarity by (24);\\
8: $~~~$ Use the learned model from the expert; \\
9: $~~~$ Update the overall Q-table by (25); \\
10: $~~$ Update the transfer rate $\mu $, and select an action by (31); \\
11: $~~$ Perform learning from step 13 to step 16;\\
12: $~$  \textbf{else} \\
$~~~~$ \textbf{\emph{Cooperative learning}}\\
13: $~~$ In each group $g$, each agent shares its observations and actions;\\
14: $~~$ Each group calculate its cooperative Q-table by (28); \\
15: $~~$ Update the joint policy ${\pi _g}{\rm{(}}{s_g}{\rm{)}}$ with the largest cooperative Q-value ${Q_g}({s_g},{a_g})$, and  selecte the joint action ${a_g}$; \\
16: $~~$ Execute action ${a_g}$, then obtain a reward  ${r'_g}$ using (11), and observe a new state ${s'_g}$;\\
17: $~$ \textbf{end if}\\
18: $~$  Both transfer learning and cooperative learning are jointly updated to optimize the learned policy;\\
19: \textbf{end loop}\\
20:  \textbf{output:} Subchannel assignment and power control. \\
\end{small}
\label{alg_lirnn}
\end{algorithm}

Multi-agent reinforcement learning is also called "independent DQN", where each agent independently learns its own policy and considers other agents as part of the environment. Moreover, , the combination of experience replay with independent DQN appears to be problematic: the non-stationarity introduced by independent DQN. Hence, we have presented a distributed cooperative multi-agent DQN scheme, devices are encouraged to communicate and share their learned experiences and actions within a small number of neighbors, and finally learn with each other. In this case, the scheme is capable of avoiding the non-stationarity of independent Q-learning by having each agent learn a policy that conditions on an information-sharing of the other agents’ policies (behaviors) in the same group.

\subsection{Computational Complexity Analysis}

For the training phase, in trained DQN models, let $L$,
$B_0$ and $B_l$ denote the training layers which are
proportional to the number of states, the size of the input layer
and the number of neurons used in DQN, respectively. The
complexity in each time step for each agent is calculated by
$O(B_0B_1 + \sum\nolimits_{l = 1}^{L - 1}
{B_lB_{l + 1}} )$ at each training step. In the
training phase, each mini-batch has episodes ${N^{{\rm{epi}}}}$
with each episode being $T$ time steps, and each trained model is
completed over $I$ iterations until convergence and the network
has $Z$ agents with the $Z$ trained DQN models. Hence, the total
computational complexity is $O\left(
{ZI{N^{{\rm{epi}}}}T(B_0B_1 + \sum\nolimits_{l = 1}^{L
- 1} {B_lB_{l + 1}} )} \right)$. The high
computational complexity of the DQN training phase can be
performed offline for a finite number of episodes at a powerful unit (such as the BS) [38], [39].

For the distributed cooperative phase (also called testing phase),
our proposed approach applies the transfer learning mechanism and allows
the expert agent to share the learned knowledge or actions  with other agents. Let ${\mathcal{S}}'$
and ${\mathcal{A}}'$ denote the stored state space and action space, respectively. The computational
complexity of the classical DQN approach (the fully distributed
DQN approach) and the proposed approach are $O(|S{|^2} \times
|{\mathcal{A}}|)$  and $O(|{\mathcal{S}}'{|^2} \times |{\mathcal{A}}'| + |{\mathcal{S}}{|^2} \times |{\mathcal{A}}|)$ [19],
respectively, indicating that the complexity of the proposed
approach is higher than the classical DQN
learning approach. Nevertheless, the stored state space and action space in the memory is
not large at each device, and  hence the complexity of the proposed learning approach is slightly higher
than the classical DQN approach. For cooperative learning, a small number of agents in each same group will select their actions jointly instead of independently by sharing their own selected action. Let  $a^{co}_{g,i}$ denote the shared action set of each $i$-th agent in the $g$-th group in the current time slot, then the computational complexity the $g$-th group in term of action sharing is $O(\sum\nolimits_{i = 1}^{{L_g}}
{|a^{co}_{g,i}|})$. As the network has $G$ groups, the total computational complexity of the cooperative learning is $O(\sum\nolimits_{g = 1}^{{G}} \sum\nolimits_{i = 1}^{{L_g}}
{|a^{co}_{g,i}|})$.

\section{Simulation Results and Analysis}

In this section, simulation results are provided to evaluate the
proposed distributed cooperative multi-agent RL based massive
access approach. We consider a single cell with a cell radius of
500 m, the total number of devices is 2000.
In addition, we set one fifth of the total number of devices to be
normal services and the minimum data rate requirement is set as 3.5 bps/Hz.
The maximum D2D communication distance is 75 m. The carrier frequency is 2 GHz, and the total bandwidth
is 100 MHz which is equally divided into 100 subchannels with each
subchannel having  1 MHz. For the URLLC services, the SINR
threshold is 5 dB, the processing/computing delay
${T_{{\rm{pc}}}}$= 0.3 ms, the reliability requirement varies
between 99.9\% and 99.99999\%, and the maximum latency threshold
varies between 1 ms and 10 ms for different simulation settings.
The maximum transmit power of each device and circuit power
consumption are 500 mW and 50 mW, respectively. The background
noise power is -114 dBm. Each packet size in URLLC links is 1024
bytes. The DQN model consists of three connected hidden
layers, containing 250, 250, and 100 neurons, respectively. The
learning rate is $\alpha  = 0.02$ and discount factor is set to be
$\gamma  = 0.95$.

\begin{table}[!t]
\renewcommand{\arraystretch}{1.0}
\caption{Simulation Parameters} \centering
\includegraphics[width=0.475\textwidth]{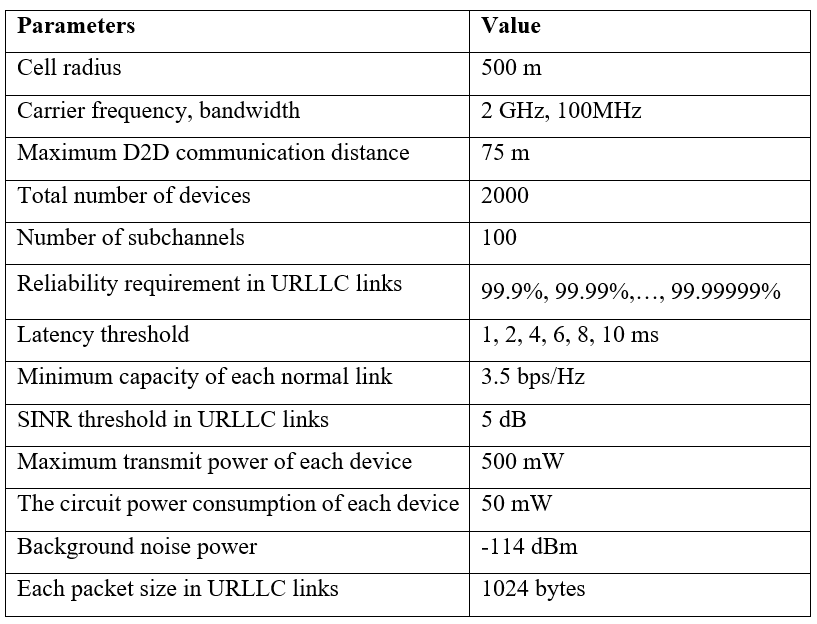}
\end{table}

We compare the proposed distributed cooperative multi-agent RL
based massive access approach (denoted as proposed DC-DRL MA, which adopts both   transfer learning and cooperative learning mechanisms) with
the following approaches:

\emph{1)} The group based massive access approach, where devices
are grouped by the similarities with each group having one group
leader to communicate with the centralized controller. Then, the
subchannel assignment and transmission power control are adjusted
iteratively to the communication links in each group, similar to
the group based preamble reservation access approach [13] (denoted
as centralized G-MA).

\emph{2)} The fully distributed multi-agent RL based massive
access approach (denoted as fully D-DRL MA [37]), similar to the
approach [37], where each communication link selects its
subchannel assignment and transmission power strategy based on its
own local information without cooperating with other communication
links.

\emph{3)} Random massive access approach (denoted as random MA),
where each communication link chooses its subchannel assignment
and transmission power strategy in a random manner.

\subsection{Convergence Comparisons}

Here, we show in Fig. 4 the energy efficiency (EE) with increasing training episodes to investigate the convergence behavior of the proposed multi-agent DQN approach and compared approaches. Clearly, the proposed learning approach significantly achieves the higher EE performance  than that of the fully distributed DRL approach [37] and random MA approach. Especially, the proposed approach has faster convergence speed and less fluctuations by adopting by transfer learning and cooperative learning mechanisms to improve the learning efficiency and convergence speed. The fully distributed DRL approach [37] is simple without any cooperation among devices, but it achieves poor global performance, leading to the poor EE value. Even though the random MA approach has the simplest structure, the worst performance fails to optimize the network energy efficiency with increasing training episodes. Our proposed approach applies both the transfer learning and cooperative learning mechanisms to enhance the convergence speed and learning efficiency, and the optimized strategy can be learned after a number of training episodes. From Fig. 4, the energy efficiency per episode improves as training continues, demonstrating the effectiveness of the proposed training approach. When the training episode approximately reaches 1900, the performance gradually converges despite some fluctuations due to mobility-induced channel fading in mobile environments. Since we investigate the resource management in massive access scenario, the environment is complex, as well as the action and state spaces are large for all mobile devices, so our presented learning approach requires about 2000 training episodes to appropriately converge. 

\begin{figure}
\centering
\includegraphics[width=0.85\columnwidth]{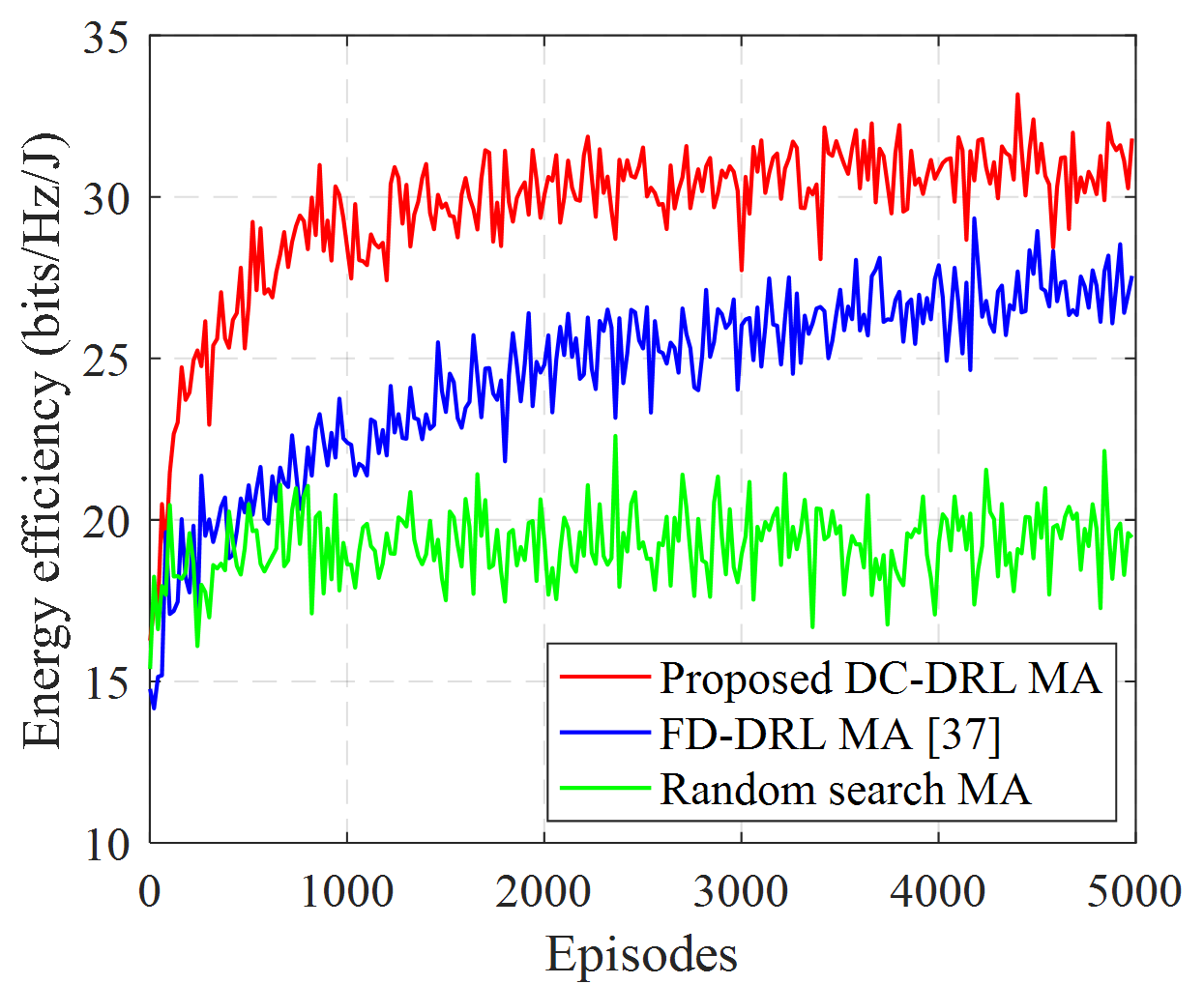}
\caption{{Convergence comparisons of compared learning approaches.   } } \label{fig:Schematic}
\end{figure}

\subsection{Performance Comparisons Under Different Thresholds of
Reliability and Latency}

Fig. 5 and Fig. 6 compare the performances of all approaches under
different values of the reliability and latency thresholds,
respectively, when the packet arrival rate is 0.03
packets/slot/per link and the total number of devices is 2000.
From both Fig. 5 and Fig. 6, for all approaches, we can find that
both the EE performance and the transmission success probability
drop as the required reliability value increases and the maximum
latency threshold decreases. The reason is that the more
stringent the reliability and latency constraints are, the worse
network EE and transmission success probability the network can
archive. In this case, both the transmission power and subchannel
assignment strategy needs to be carefully designed to guarantee
the stringent reliability and latency constraints, such that the
transmission success probability can be guaranteed at a high
level.

 \vspace{-2pt}
\begin{figure}
\centering
\includegraphics[width=1\columnwidth]{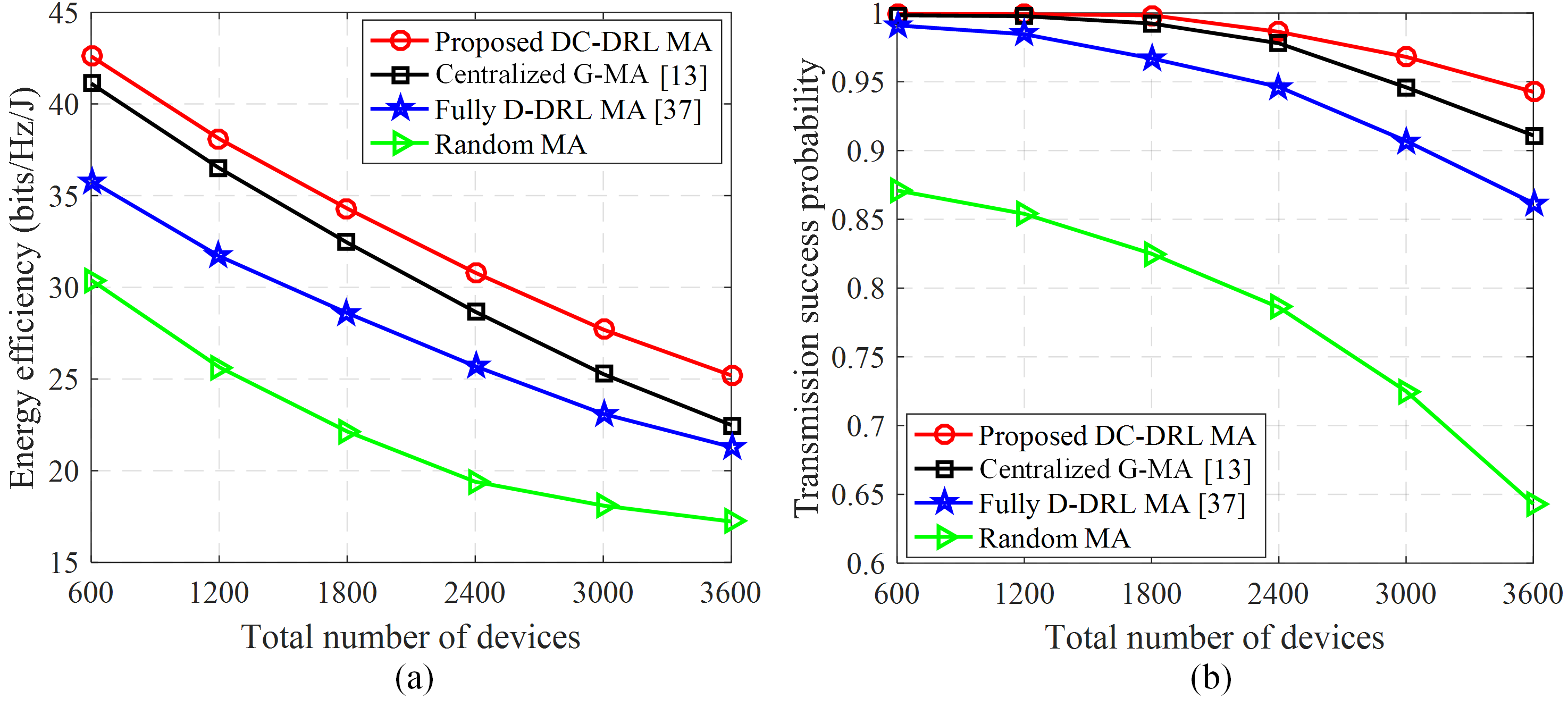}
\vspace{-2pt} \caption{{\small Performance comparisons vs.
different reliability thresholds.} } \label{fig:Schematic}
\vspace{-5pt}
\end{figure}

\vspace{-2pt}
\begin{figure}
\centering
\includegraphics[width=1\columnwidth]{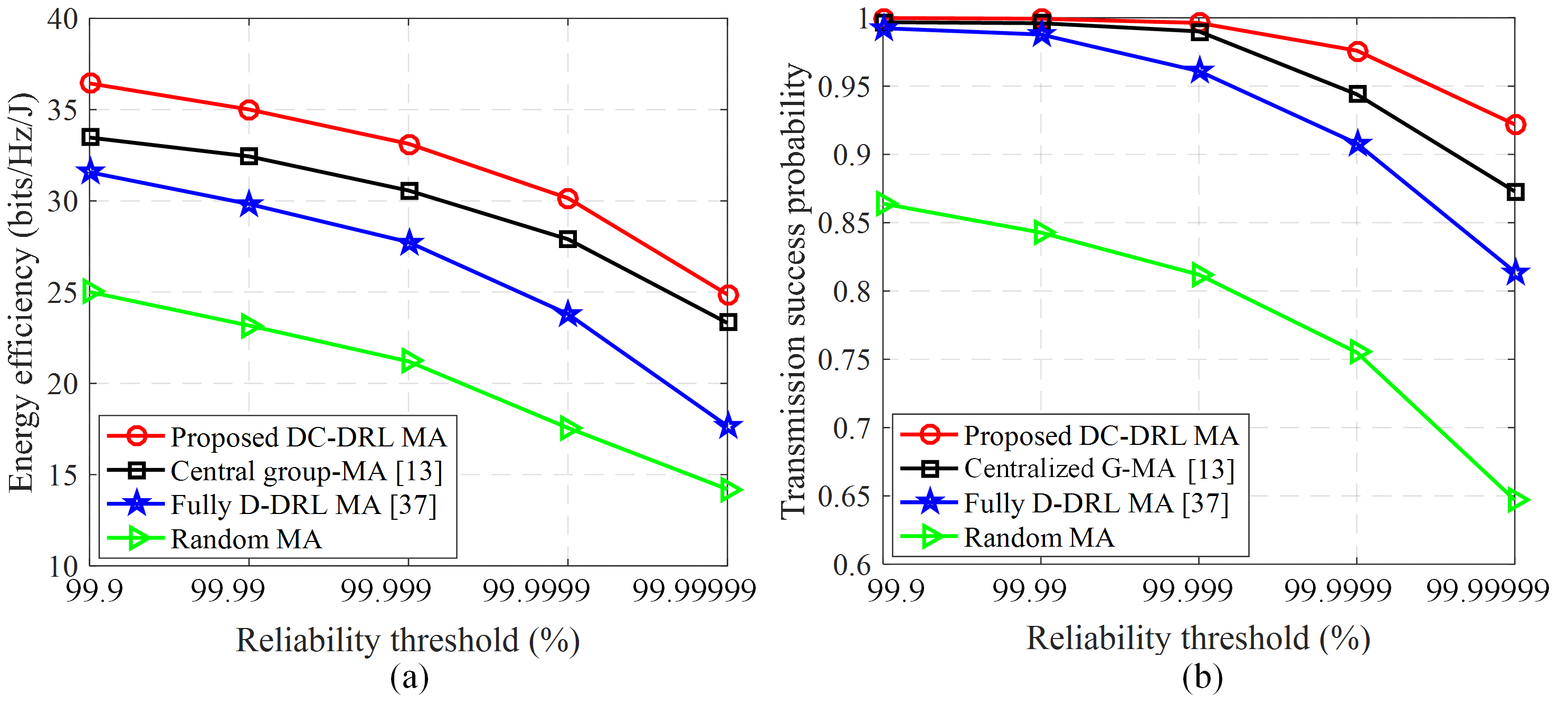}
\vspace{-2pt} \caption{{\small Performance comparisons vs.
different latency thresholds.} } \label{fig:Schematic}
\vspace{-5pt}
\end{figure}

We also observe from Fig. 5 (b) and Fig. 6 (b) that within a
reasonable region of the reliability and latency thresholds
change, the three approaches (except the random search approach)
can till achieve the high transmission success probability, which,
however, have more unsatisfied transmission link events happen if
the constraints are extremely stick (e.g., the reliability
threshold grows beyond 99.999\% or the maximum latency threshold
is less than 4 ms). Compared with other approaches, our proposed
approach achieves the higher EE performance and transmission
success probability under different reliability and latency
requirements, especially the performance gap between the proposed
approach and other approaches becomes more significant when the
constraints become more stringent. The reason is that our
proposed approach employs both the transfer learning and
cooperative learning mechanisms to optimize the global subchannel
assignment and transmission power strategy, thereby improving  the
network performance. From Fig. 5 (a) and Fig. 6 (a),
an interesting observation is that compared with the centralized G-MA
approach [13] and random MA approach, the EE value curve declines
more quickly in our proposed approach when the constraints become
stricter. The reason is that the proposed approach designs the
specific QoS-aware reward function shown in (15) to try to
guarantee QoS requirements (meeting the high transmission success
probability), and hence  the network may sacrifice the part of EE
performance to support more successful transmission communication
links.

\subsection{Performance Comparisons Versus Packet
Arrival Rate}

Fig. 7 represents the performance comparisons with respect to the
increasing packet arrival rate   for different massive access
approaches, when the number of devices is 2000, and the
reliability and latency thresholds are 99.999\% and 5ms,
respectively. From Fig. 7, with the growing packet arrival rate, both
the EE performance and the transmission success probability
decrease slightly for all approaches when the packet arrival rate
is less than a certain threshold, but drop sharply when the
packet arrival rate grows beyond the acceptable margin. An
increase of packet arrival rate results in longer transmission
duration (e.g., transmission delay and queue waiting delay),
frequent spectrum access and possibly increases more transmission
power in order to improve packet transmission success probability.
In addition, the increase of packet arrival rate also leads to
stronger co-channel interference for a longer period, which limits
the data rate improvement. Hence, as shown in Fig. 7 (a), the EE
performance decreases slightly with the increase of packet
arrival rate when packet arrival rate is not high, and the
performance will become worse if packet arrival rate grows beyond
the acceptable margin.

\vspace{-2pt}
\begin{figure}
\centering
\includegraphics[width=1\columnwidth]{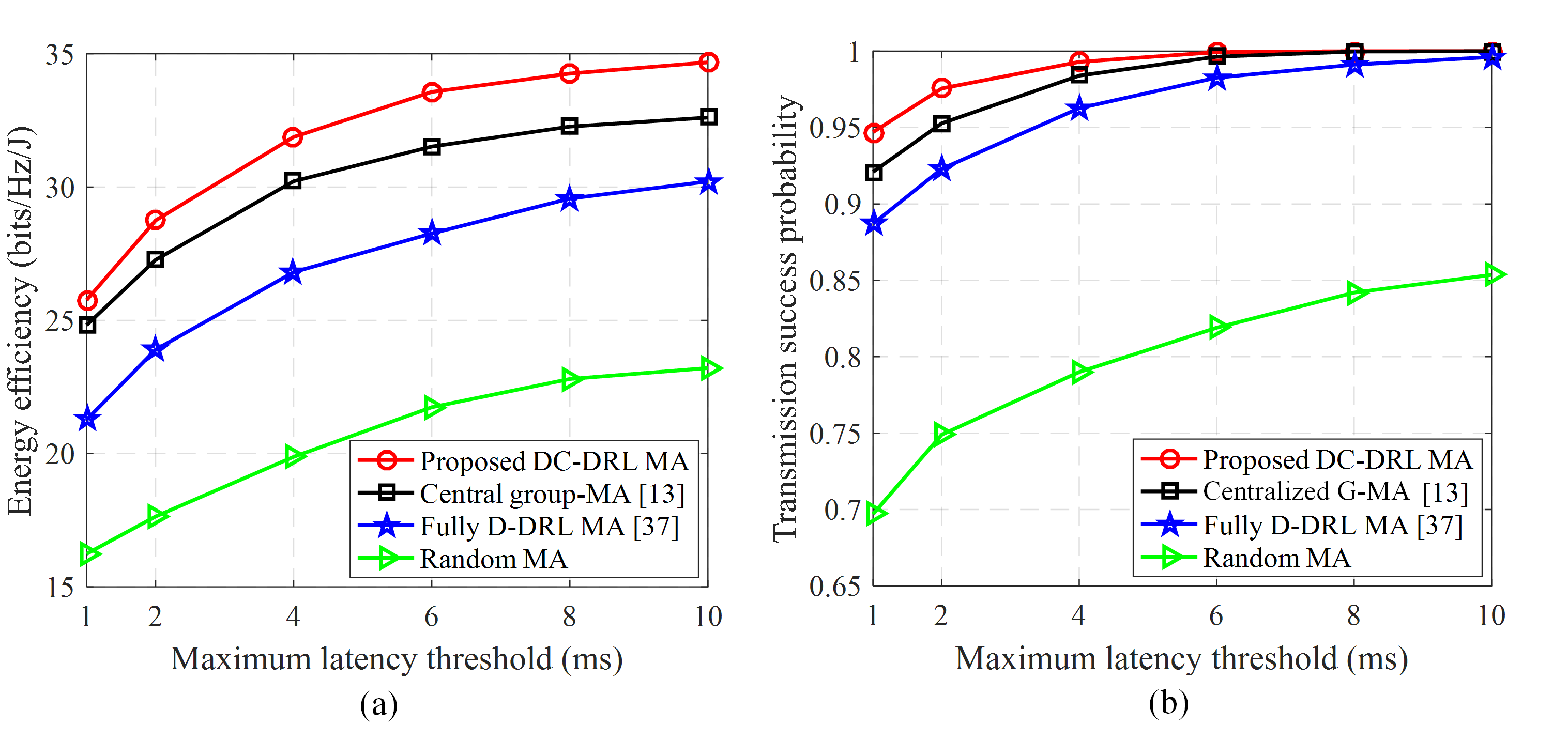}
\vspace{-2pt} \caption{{\small Performance comparisons with
different packet arrival rates.} } \label{fig:Schematic}
\vspace{-5pt}
\end{figure}

From Fig. 7 (b), even though the transmission success probability
drops for all approaches, the proposed approach still achieves  the
better performance than other three approaches. Remarkably, the
proposed approach attains approximately 100\% success transmission
probability when the packet arrival rate is less than 0.03 packets
per time slot, and achieves noticeable degradation when the packet
size grows beyond 0.03 packets per time slot. Such the performance
degradation may result from the limited spectrum resource, where
the current subchannel resource cannot completely support the
massive number of transmission packets simultaneously with the
increasing packet arrival rate.

\section{Conclusion}

In this paper, a distributed cooperative channel assignment and
power control approach based on multi-agent RL has been presented
to solve the massive access management problem in future
wireless networks, where the proposed approach is capable of
supporting different QoS requirements (e.g., URLLC and minimum data rate) of a huge number of devices. The
proposed multi-agent RL based approach consists of a centralized
training procedure and a distributed cooperative implementation
procedure. In order to improve the network performance and QoS
satisfaction levels, the transfer learning and cooperative
learning mechanisms have been employed to  enable communication links to work cooperatively in a
distributed cooperative way. Simulation results have confirmed the
effectiveness of the proposed learning approach and also showed
that the proposed approach outperforms other existing approaches
in massive access scenario.

\end{document}